\documentstyle[11pt]{article}
\begin{document}
%\begin{titlepage}

\vspace{0.5cm}

\centerline{\Large \bf The $K^0$ condensed phase of color-flavor
locked quark matter :}  \centerline{\Large\bf its formation as a
stable state reexamined}
\vspace{1cm} \centerline{ Xiao-Bing Zhang
and Xue-Qian Li } \vspace{0.5cm} \centerline{\small Department of
Physics, Nankai University, Tianjin 300071, China} \vspace{8pt}

\vspace{1cm}

\begin{minipage}{13cm}
\centerline{\bf Abstract} {\rm\noindent Taking the vector
kaon-quark interaction into account, we reexamine in detail the
neutral kaon condensation in color-flavor locked quark matter. It
is found that the pairing phenomena in the quark matter are
somewhat influenced by $K^0$ condensation. Different from the
previous predictions, we find that the $K^0$ condensed phase might
no longer be more stable than the normal phase even if $K^0$
condensation can occur in the color-flavor locked matter.}

\vspace{0.5cm} {\bf PACS number(s): 12.39.Fe, 11.30.Rd}
\date{today}

\end{minipage}

\baselineskip 18pt

\vspace{1.0cm} \noindent {\bf I. Introduction} \vspace{0.5cm}

Investigations of strongly interacting matter at high baryon
density have attracted much attention for years. In recent years,
the studies of the physics with dense quark matter become a "hot"
topic since some color superconducting phases were proposed for
high-density QCD. For the case of three massless flavors, it has
been suggested that the original color and flavor
$SU(3)_{color}\times SU(3)_{L} \times SU(3)_{R}$ symmetries of QCD
are broken down to the diagonal subgroup $SU(3)_{color+L+R}$ via
BCS like pairing \cite{alf}. This symmetry breaking pattern
exhibits a resemblance to the ordinary chiral symmetry breaking in
the low-density QCD case. Since color and flavor rotations are
locked together, such a particular symmetric state is called the
color-flavor locked (CFL) phase. It is widely accepted to be a
possible ground state of the dense quark matter with three
flavors. The main low-energy degree of freedom in this phase is an
octet of pseudo Goldstone modes that are associated with the
spontaneous chiral symmetry breaking. The effective Lagrangian
describing the Goldstone modes in dense quark matter was studied
in Refs.\cite{cas,hong,son}. Recently, condensation of the
Goldstone modes has been investigated by using the chiral
effective Lagrangian \cite{schrl,sch,kr,red}. Bedaque and
Sch\"{a}fer indicated that the stress induced by the strange quark
mass $m_s$ triggers the condensation of neutral kaon modes
\cite{sch}. The $K^0$-mode condensed phase of CFL quark matter,
which is called as the CFL$K^0$ phase in the following, was
believed to be more realistic than the normal CFL phase without
condensations \cite{sch,kr}. Therefore, the CFL$K^0$ phase is
expected to exist in core of neutron stars and investigations on
this phase are relevant to the physics of compact stars
\cite{kr,red}. The formation of $K^0$-mode condensation makes it
possible that CFL quark matter reduces its strange content. The
hadronic matter at finite density possesses non-zero strangeness
because of the presence of hyperons and/or the kaon-meson
condensation. From this point of view, investigation on the
kaon-mode condensed phases of CFL matter might be necessary for
exploring the QCD phase diagram such as the phase transition from
nuclear matter to CFL quark matter \cite{sw,alfd,zhang2}.

In the references cited above, the properties of CFL matter is
assumed to be unchanged regardless of whether $K^0$ condensation
occurs or not. Thus the CFL$K^0$ phase can be understood as a
simple " mixture " of normal CFL matter and the system consisting
of the condensed kaons, namely CFL$K^0$=CFL+ $K^0$.
Correspondingly, the phase transition between the CFL and CFL$K^0$
phases is of second order \cite{sch,kr,red}. Based on this
assumption, it seems to be an undoubted conclusion that the
CFL$K^0$ phase is \emph{always} more stable than the CFL phase.
Nevertheless, some details concerning the stability of the
CFL$K^0$ phase need to be further investigated. Generally the
CFL$K^0$ phase should be different from the normal CFL phase in
all ways. Therefore there might exist some effects of $K^0$
condensation on the CFL matter in which the condensed kaons
emerge. Such effects are expected to manifest once the $K^0$
condensation with finite condensate strength has appeared.

On the other hand, it is found more recently that the gapless
phases of CFL matter might replace the ordinary CFL phases where
there is a common pairing gap denoted as $\Delta$ \cite{alf03}. If
the electron chemical potential is ignored, the pairing between
the green-strange and blue-down quarks might become unstable
gradually with the increasing strange quark mass $m_s$
\cite{alf03}. Thus, the corresponding gapless phase replaces the
CFL/CFL$K^0$ phase as long as $m_s$ is large enough. Noticing that
$K^0$ condensation is caused by the nonzero value of $m_s$
essentially, the question arises naturally whether and how the
CFL$K^0$ phase exists as a stable state if $m_s$ is relatively
large. If the assumption of CFL$K^0$=CFL+ $K^0$ were correct so
that the larger $m_s$ is, the more stable the CFL$K^0$ phase would
be \cite{sch,kr}. Then, the CFL$K^0$ phase would be energetically
favorable eventually and even be favored over the gapless phase.
Obviously, the above conjecture is contradictory to the existence
of the gapless phases. In this sense, the previous treatment for
the CFL$K^0$ phase on the basis of CFL$K^0$=CFL+ $K^0$ might be
problematic. This discrepancy motivates us to reexamine the
mechanism of $K^0$ condensation and the stability condition for
the CFL$K^0$ phase. Taking an effective kaon-quark interaction
into account, we find that the CFL pairing phenomena in the quark
matter are influenced by $K^0$ condensation. As a consequence, the
CFL$K^0$ phase might be no longer energetically favorable with
respect to the CFL phase even if $K^0$ condensation can occur.
Moreover, we suggest that the transition from the CFL phase to the
CFL$K^0$ phase is generally a first order transition, instead of
the second order one.

\vspace{0.5cm} \noindent {\bf II. $K^0$ condensation in CFL
matter} \vspace{0.5cm}

Let us begin with the description of CFL quark matter when no
condensation occurs. The pressure of the CFL phase can be written
as
\begin{eqnarray} {\cal P}_{CFL}= {\cal P}_{unpair} +{\cal P}_{pair},\label{pcfl}
\end{eqnarray}
where ${\cal P}_{unpair}$ denotes the pressure from the unpaired
quark matter where all the three flavors have the same Fermi
momenta and its form is model dependent. The pressure induced by
the CFL pairing is given by \cite{alfd,raj}
\begin{eqnarray} {\cal P}_{pair}=\frac{3\Delta^2
\mu^2}{\pi^2},\label{ppair}
\end{eqnarray}
where $\mu$ is the quark chemical potential and $\Delta$ is the
CFL pairing gap in the color triplet. The baryon density of the
CFL phase reads \cite{alfd}
\begin{eqnarray} {\rho}_{CFL}&=& \frac{p_F^3+2\Delta^2 \mu}{\pi^2},\label{ncfl}
\end{eqnarray}
and the common Fermi momentum $p_F$ is required to be \cite{raj}
\begin{eqnarray}
p_F=2\mu-\sqrt{\mu^2+m_s^2/3}\approx\mu-m_s^2/(6\mu),\label{pf}
\end{eqnarray}
in order to guarantee the electrical neutrality of the CFL quark
matter.

For the Goldstone modes in CFL quark matter, the leading terms of
the concerned effective Lagrangian are given as \cite{cas,sch}
\begin{eqnarray}
{\cal L}_{G}&=& \frac{f_\pi^2}{4}Tr(\partial_0 U
\partial_0 {U^+}- v^2_\pi \partial_i U
\partial_i U^+)-A[det(M)Tr(M^{-1}U) +
h.c.], \label{lg}
\end{eqnarray}
where $v_\pi$ is the 3-velocity of the Goldstone modes, $M$ is the
quark mass matrix and the chiral field $U$ is defined by
$U=\exp(i\sqrt{2}\Phi/{f_\pi})$ with the octet of Goldstone modes
$\Phi$. By perturbative calculations at high density QCD, the pion
decay constant $f_\pi$ and the low energy coefficient $A$ are
\cite{son}
\begin{eqnarray} {f_\pi^2}= \frac{21-8{\ln 2}}{18}
\frac{\mu^2}{2\pi^2},\label{fpi}
\end{eqnarray}
and
\begin{eqnarray}
{A}= \frac{3\Delta^2}{2\pi^2},\label{a}
\end{eqnarray}
respectively. By Eqs.(\ref{lg}) and (\ref{a}), the masses of
Goldstone modes are given by
\begin{eqnarray}
{m_\pi^2}=\frac{6\Delta^2}{\pi^2 f_\pi^2}m_q m_s ,\,\,\,\,
\,{m_K^2}=\frac{3\Delta^2}{\pi^2 f_\pi^2}m_q(m_q+m_s), \label{mk}
\end{eqnarray} where $m_q=m_{u,d}$ is the light quark mass in the
limit of isospin symmetry. Eq.(\ref{mk}) shows that the kaon-mode
is lighter than the pion-mode\footnotemark[1] \footnotetext[1]{
The anomalous mass ordering is closely linked to the Lagrangian
Eq.(\ref{lg}), where the Goldstone-mode mass term contains
$det(M)M^{-1}$. If other possible terms, say the instanton
interaction, were considered, this behavior would be modified. In
that case, the condensations of Goldstone-modes might not occur
\cite{hong,bs}. In the present work we assume that the instanton
contribution to the Goldstone-mode mass is negligible at least for
large quark chemical potential. }, which implies that the
formation of kaon condensation is easier than the pion
condensation in the CFL environment.

In the vicinity of the Fermi surface, quasi-quark excitations are
active and thus become the natural degrees of freedom in the CFL
effective theory. The interaction among these quasi quarks has
been considered in the framework of high-density effective theory
\cite{hong,sch,bs}. The particle excitation can be described by
$\tilde{\psi}$ defined as
$\tilde{\psi}=\frac{1}{2}(1+\vec{\alpha}\cdot\hat{v}_F)\psi$,
where $\psi$ is the quark field and $\hat{v}_F$ is the Fermi
velocity, in the effective theory \cite{sch}. Using the
$\tilde{\psi}$ field, such a term as $
-\frac{1}{\mu}(\tilde{\psi_L}^+ MM^+\tilde{\psi_L} +
\tilde{\psi_R}^+ M^+M\tilde{\psi_R})$ is introduced at the leading
order in $1/{\mu}$ \cite{hong,sch}. In the case of $m_s \gg m_q$,
this term is further simplified as
\begin{eqnarray}
{\cal L}_{q} = -\frac{m_s^2}{2\mu} \tilde{\psi^s}^+ \tilde{\psi^s}
.\label{lq2}
\end{eqnarray}
In Ref.\cite{sch}, Bedaque and Sch\"{a}fer regarded
$\frac{m_s^2}{2\mu}$ in Eq.(\ref{lq2}) as the chemical potential
for the strange hole excitations. According to the chemical
equilibrium, the chemical potential for $K^0$ modes is
\begin{eqnarray} \mu_K = \frac{m_s^2}{2\mu}, \label{muk}
\end{eqnarray}
and thus the critical condition of $K^0$ condensation is required
to be $\mu_K= m_K$. As argued in Refs.\cite{sch,kr}, the CFL phase
was predicted to undergo a second order transition to the CFL$K^0$
phase as long as $m_s^2/(2\mu)$ is larger than $m_K$ or
approximately
\begin{eqnarray}
m_s \geq (\frac{2\sqrt{3}\mu}{\pi f_\pi})^{2\over 3} m_q^{1\over
3} \Delta^{2\over 3}\approx 3 m_q^{1\over 3} \Delta^{2\over 3}.
\label{cri}
\end{eqnarray}

It is well known that the condensate strength is closely linked to
the ratio of the kaon chemical potential and the kaon mass. With
the increasing value of $m_s^2/(2\mu)$, both the density of the
condensed kaons and the pressure from kaon condensation arise
rapidly. For instance, if $m_s^2/(2\mu)$ is close to $\Delta$, the
condensate density has an order of $\Delta \mu^2/(2\pi^2)$ and the
condensate pressure has an order of $\Delta^2 \mu^2/(2\pi^2)$.
Comparing with Eqs.(\ref{ppair}) and (\ref{ncfl}), they are of the
same orders as the pairing contributions to the pressure and the
baryon density in the CFL quark matter. In this case, the $K^0$
condensation should be \emph{not merely} caused by the quasi
quarks in the vicinity of the Fermi surface, which have taken part
in the CFL pairing. In order to make the $K^0$ condensation with
large condensate strength possible, it is reasonable to assume
that an amount of quasi quarks residing away from the Fermi
surface also take part in the process
$\tilde{\psi^s}^+\tilde{\psi^d}\rightarrow K^0$. Now that more
quasi quarks than those near the given Fermi surface get involved
in the $K^0$ condensation and the kaon chemical potential not only
originates from the non-zero strange quark mass, but also comes
from the contribution induced by these additional quasi quarks. In
other words, the kaon chemical potential is expected to be
deviated from the value given by Eq.(\ref{muk}) when the
condensate strength is large, as will be explained in the next
section.

\vspace{0.5cm} \noindent {\bf III. Vector kaon-quark interaction
and its effect on the CFL$K^0$ phase} \vspace{0.5cm}

According to Refs.\cite{sch,kr}, the $K^0$ condensation is caused
by the quark self-interacting term Eq.(\ref{lq2}) exclusively.
This treatment is very different from the kaon-meson condensation
in hadronic environment such as in nuclear matter. As pointed out
in Refs.\cite{kn}, the effective interaction between baryons and
kaons plays the key role in meson condensation. In the CFL
environment, the quasi quarks have an energy gap $\Delta$ so that
the interaction between kaon-modes and quarks is usually
considered to be very small. Even if so, the effective kaon-quark
interaction needs to be included in the case that the $K^0$
condensate strength is large enough. There are many types of
kaon-quark interactions such as scalar coupling, vector coupling,
pseudoscalar coupling and pseudovector coupling. For the purpose
of this work \footnotemark[2] \footnotetext[2]{ We have considered
the scalar interaction caused by the non-zero strange quark mass
in CFL matter \cite{zhang}. Since its effect on $K^0$ condensation
is found to be small, we will ignore it in this work. As for the
pseudoscalar and pseudovector interacting terms, their effects can
be neglected at the mean-field level as well as in the limit of
isospin symmetry .}, our main intention is to consider the leading
term in the vector kaon-quark interactions
 \begin{eqnarray} {\cal
L}_{kq}=-\frac{i C}{ f_\pi^2} \tilde{\psi}^+
\tilde{\psi}(\overline{K} \overrightarrow{\partial_0} K -
\overline{K} \overleftarrow{\partial_0} K) , \label{lkq}
\end{eqnarray}
where $C$ is the coupling coefficient. Eq.(\ref{lkq}) bears a
formal analogy with the Tomozawa-Weinberg term accounting for the
leading vector interaction between nucleons and kaon-mesons
\cite{kn,mish}. Both Eq.(\ref{lkq}) and the Tomozawa-Weinberg term
originate from the vector coupling $\sim \overline{B}\gamma^\mu
v_\mu B$, where ${v}_{\mu}\equiv \frac{i}{2}(\xi\partial_\mu
{\xi^+} +{\xi^+}\partial_\mu \xi)$ is the vector current of
Goldstone bosons with $\xi=\sqrt{U}=\exp(i\Phi/\sqrt{2}{f_\pi})$
and $B$ denotes the baryon/quark field. Since we only concern the
kaon mode in the Goldstone matrix, it is practical to consider the
form of Eq.(\ref{lkq}) at the mean field level. In addition,
Eq.(\ref{lkq}) is valid even in the chiral limit and it is
responsible for the dynamical breaking of chiral symmetry.
Noticing that Eq.(\ref{lq2}) is responsible for the explicit
chiral breaking, thus incorporation of Eqs.(\ref{lq2}) and
(\ref{lkq}) is necessary for reflecting the symmetry breaking
pattern in the CFL context.

When $K^0$ condensation occurs, the expectation values of the
neutral kaon and anti-kaon fields can be expressed as
\cite{sch,kn}
\begin{eqnarray} \langle{K^0}\rangle =\frac{f_\pi}{\sqrt{2}}\theta \exp(-i\mu_{K} t),
\,\,\,\,\,\,  \, \langle{{\overline{K}^0}\rangle
=\frac{f_\pi}{\sqrt{2}}\theta \exp(i\mu_{K} t)}  ,
\label{kco}\end{eqnarray} respectively, where $\theta$ is a
dimensionless parameter determining the condensate strength. It is
easy to find that the expectation value of ${i(\overline{K}
\overrightarrow{\partial_0} K - \overline{K}
\overleftarrow{\partial_0} K)}$ can be replaced by the condensate
density denoted as $\rho_{con}$ and then Eq.(\ref{lkq}) becomes
\begin{eqnarray}
-\frac{C\rho_{con}}{ f^2_\pi} \tilde{\psi}^+ \tilde{\psi},
\label{lkq2}
\end{eqnarray}
at the mean field level. Comparing Eq.(\ref{lkq2}) with
Eq.(\ref{lq2}), it means that ${C\rho_{con}}/( f^2_\pi)$ can be
regarded as an additional effective chemical potential for the
strange hole excitations. As expected, there is a shift in the
kaon chemical potential due to the presence of $K^0$ condensation.

To treat Eq.(\ref{lkq2}) self-consistently, we need to replace
$\mu_K$ by ${\mu_K}'$ as
\begin{eqnarray} \mu_K \rightarrow {\mu_K}'=
\frac{m_s^2}{2\mu}+\frac{C\rho_{con}}{ f^2_\pi}, \label{muk2}
\end{eqnarray}
into the calculations for $K^0$ condensation. Then, the condensate
density and the condensate pressure take the forms :
\begin{eqnarray}
\rho_{con}= {f_\pi^2}{\mu_K}'
 [1-\frac{m_{K}^4}{({\mu_K}')^4}],\label{qq2}
\end{eqnarray}
and
\begin{eqnarray}
{\cal P}_{con}= \frac{1}{2}{f_\pi^2}({\mu_{K}}')^2
 [1-2\frac{m_{K}^2}{({\mu_{K}}')^2}+\frac{m_{K}^4}{({\mu_{K}}')^4}],
\label{pcon} \end{eqnarray} respectively, instead of those given
previously \cite{kr,alfd}. Eqs.(\ref{qq2}) and (\ref{pcon}) are
valid only for the case that $K^0$ have condensed.

This is not the whole story yet. The quasi quarks residing away
from the Fermi surface could not become active unless they gain
pressure from that induced by the CFL pairing. Therefore, the
$K^0$ condensation should affect the pairing phenomena in the CFL
matter in turn. As pointed out by Alford \emph{et. al.}
\cite{alf03}, the CFL pairing gap is unchanged approximately as
long as $m_s^2/(2\mu)$ is still smaller than $\Delta$. In this
case, the $K^0$ condensation can affect the pairing pressure via
changing the magnitude of $\mu$ in Eq.(\ref{ppair}) equivalently.
In view of the fact that the shift in the kaon chemical potential
Eq.(\ref{muk2}) is a sort of Fermi energies essentially, we simply
assume that the " effective " quark chemical potential is
suppressed by the same shift, namely $\frac{C\rho_{con}}{
f^2_\pi}$. Here we stress that the actual quark chemical potential
is still determined by the baryon density of the unpaired quark
matter, so that the definition of Eq.(\ref{pf}) is not influenced
by the $K^0$ condensation. In other words, the decrease of the "
effective " quark chemical potential makes sense only for the
pairing phenomena in the CFL matter. Therefore, the pairing
pressure Eq.(\ref{ppair}) and then the baryon density
Eq.(\ref{ncfl}) are replaced by
\begin{eqnarray}
{\cal P}_{pair}=\frac{3\Delta^2 {(\mu-\frac{C\rho_{con}}{
f^2_\pi})}^2}{\pi^2},\label{ppair2}
\end{eqnarray}
and
\begin{eqnarray} {\rho}_{CFLK^0}&=& \frac{p_F^3+2\Delta^2
(\mu-\frac{C\rho_{con}}{ f^2_\pi})}{\pi^2},\label{ncflk}
\end{eqnarray}
respectively.

Eq.(\ref{ncflk}) is the baryon density of the CFL matter in which
kaons condense. It implicates that the baryon density in the
CFL$K^0$ phase is reduced more or less, which is an indirect
result of the $K^0$ condensation effect on the CFL free energy.
This observation is different compared to the previous scenario,
where the baryon density of CFL matter is independent of whether
$K^0$ condensation occurs or not. More importantly, the pressure
in the CFL$K^0$ phase is different from the previous form. Due to
Eq.(\ref{ppair2}), the formation of $K^0$ condensation makes the
pairing pressure to be smaller than that in the CFL phase. On the
other hand, the $K^0$ condensation provides a contribution to the
pressure of the CFL$K^0$ phase. With the increasing condensate
strength, the shift in the kaon chemical potential is enlarged so
that the kaon-condensate pressure is reinforced to be larger. The
total pressure of the CFL$K^0$ phase reads
\begin{eqnarray} {\cal P}_{CFLK^0}= {\cal
P}_{unpair}+{\cal P}_{pair}+{\cal P}_{con},\label{pcflk}
\end{eqnarray}
where the second term of RHS is defined in Eq.(\ref{ppair2}).

As long as the pressure in the CFL$K^0$ phase is smaller than that
in the CFL phase, the CFL$K^0$ phase is no longer energetically
favorable. A stable CFL$K^0$ phase does not exist until the
pressures of the CFL and CFL$K^0$ phases reach a new equilibrium.
In this sense, the critical condition of the formation of the
stable CFL$K^0$ phase becomes
\begin{eqnarray}
{\cal P}_{CFLK^0}(m_s,\mu) = {\cal P}_{CFL}(m_s,\mu).\label{cri2}
\end{eqnarray}
In fact, Eq.(\ref{cri2}) is just the Gibbs condition for the first
order phase transition. Therefore, the CFL phase should undergo a
first-order transition, rather than the second-order one, to the
stable CFL$K^0$ phase.

\vspace{0.5cm} \noindent {\bf IV. Numerical results and
discussions} \vspace{0.5cm}

Finally we present some numerical examples to demonstrate how the
modified mechanism works. For sufficiently large $\mu$, the
description of the Goldstone modes in the CFL matter is valid. In
such a perturbative regime, nevertheless, the kaon condensation
effects become negligible since $m_s^2/(2\mu)$ is very small. We
may extrapolate the CFL matter from the perturbative regime to the
regime with $\mu \sim 1$GeV. For the range of $\mu=0.5-1.5$GeV, we
adopt the current estimate $\Delta\sim 100$MeV \cite{100}. Through
this work, $m_s$ is treated as a parameter whereas the quark mass
ratio is set as $m_s/{m_q}\sim20$ \cite{wein}. As for the constant
$C$ in Eq.(\ref{lkq}) we use some typical values, which are small
enough but are of order one \footnotemark[3] \footnotetext[3]{ If
the vector kaon-quark coupling is introduced in the framework of
high-density effective theory, $C$ should be of order one by power
counting. If it is introduced in the framework of the CFL
effective theory with two scales $f_\pi$ and $\Delta$, the
coupling of quark and vector current $v_\mu$ may have different
coefficients. As argued in Ref.\cite{jack}, such a coefficient
scales as $f_\pi/\Delta$ by power counting. In this case, $C$ in
Eq.(\ref{lkq}) should be still of order one for our concerned
regime. But the above estimates by the effective theory could not
be taken too seriously since our treatment of Eq.(\ref{lkq}) only
holds at the mean field level.}, in the numerical calculations. At
a given quark chemical potential $\mu= 1$GeV, Fig.1 demonstrates
the dependence of the pressures for the two phases of the CFL
matter on the strange quark mass, while Fig.2 shows the dependence
of the baryon densities on $m_s$. At $m_s\approx120$MeV,
Eq.(\ref{cri}) is satisfied and $K^0$ condensation comes to occur.
With increasing $m_s$, nevertheless, the pressure of the CFL$K^0$
phase becomes smaller than that of the CFL phase. In this case,
the $K^0$ condensed phase is energetically unstable in the CFL
matter. Also, the baryon densities in the CFL$K^0$ and CFL phases
separate from each other for $m_s>120$MeV, as shown in Fig.2. When
$m_s$ is close to $380$MeV the CFL$K^0$ and CFL phases reach a
pressure equilibrium and the CFL$K^0$ phase gradually becomes
energetically favorable as long as $m_s$ is larger than this
value. So there exists a first order CFL-CFL$K^0$ phase transition
at $m_s\approx380$MeV. Also a discontinuous jump of the baryon
density takes place at this point, as shown by the vertical dotted
line in Fig.2.

With different values of $C$, a schematic phase diagram of CFL
matter is given in Fig.3. The critical line of the CFL-CFL$K^0$
transition is obtained from Eq.(\ref{cri2}), while the boundary of
CFL matter is described by a line $m_s^2/(2\mu)=2\Delta$ at which
the CFL pairing is predicted to break down completely
\cite{alf,sch}. Also the gapless CFL phase is shown in Fig.3,
which is predicted to become dominant for the regime of
$m_s^2/(2\mu)>\Delta$ \cite{alf03}. When $C=0$, namely the
effective kaon-quark interaction is ignored, the critical line of
the CFL-CFL$K^0$ transition coincides with that given in the
literature \cite{sch,kr}. In this case, our mechanism for $K^0$
condensation comes back to the previous one, as shown by the
lowest dashed line in Fig.3. With increasing $C$, the stable
CFL$K^0$ phase is limited to a smaller window in the $(m_s,\mu)$
plane. As long as $C$ is not very small, there is not too much
room for the CFL$K^0$ phase in the $(m_s,\mu)$ plane. Depending on
the coupling coefficient of kaon-quark interaction, the $K^0$
condensed phase might be energetically disfavored compared to the
normal CFL phase. The present result is different from that drawn
by including the instanton effect, which raises the masses of
Goldstone-modes from the values given by Eq.(\ref{mk}) and thus
makes $K^0$ condensation difficult to occur \cite{red,bs}. As a
starting point, we assume that the masses of Goldstone-modes are
determined by Eq.(\ref{mk}) at least for $\mu \sim 1$GeV. Thus the
formation of $K^0$ condensation is still triggered by the non-zero
quark masses. Once the condensate density becomes finite, the
kaon-quark interaction works. Therefore the formation of the
stable CFL$K^0$ phase, i.e. the CFL-CFL$K^0$ transition, is of
first order generally. If the instanton interaction is included,
this conclusion still holds although the CFL$K^0$ phase is more
difficult to appear. For the regime with a relatively small $\mu$,
a systematic analysis on this issue needs to consider dependency
of the common $\Delta$ on $\mu$ and the color-sextet-channel
pairing too, which might change our numerical results
quantitatively.

In summary, the mechanism for $K^0$ condensation and the stability
of the CFL$K^0$ phase are examined in the phase region between the
normal CFL phase and the gapless CFL phase. Taking the vector
kaon-quark interaction into account, we find that the CFL pairing
phenomena of dense quark matter change gradually as $m_s$ is
chosen to be larger. This conclusion is essentially consistent
with the trend that the CFL pairing, say that between the
green-strange and blue-down quarks, becomes unstable with
increasing strange quark mass \cite{alf03}. Another interesting
point is that the CFL$K^0$ phase is predicted to have the lower
baryon density than the normal CFL phase. It may be important for
understanding the phenomena in the QCD phase diagram. For
instance, if the baryon density and the strangeness density in the
stable CFL$K^0$ phase become comparable with those in the
hypernuclear phase under somewhat conditions, one could not rule
out the possibility of the so-called hadron-quark continuity
\cite{sw,zhang2}. Further investigations on how the CFL$K^0$
phase, if it exists as a stable state, undergoes a transition(s)
to the gapless CFL and/or non-CFL phases need to use
model-dependent descriptions of the various CFL phases and
consider effects of the electron and color chemical potentials.
Works along these directions are worth being pursued.

\vspace{0.5cm} \noindent {\bf Acknowledgements} \vspace{0.5cm}

This work was supported by National Natural Science Foundation of
China ( NSFC ) under Contract No.10405012.

\newpage
\begin{figure}
\caption{Pressures for the CFL$K^0$ ( solid ) and CFL ( dotted )
phases as functions of strange quark mass at $\mu=1$GeV with the
choice of $C=0.5$, where the contributions from the unpaired quark
matter to pressures are subtracted. The dashed line corresponds to
the previous result, where the properties of CFL matter are not
influenced by $K^0$ condensation.}
\end{figure}

\begin{figure}
 \caption{Similar as Fig.1, but for the baryon densities. The contributions from the
unpaired quark matter are subtracted also.}
\end{figure}

\begin{figure}
 \caption{ Schematic phase diagram of CFL matter.
The dashed lines from up to bottom indicate the CFL-CFL$K^0$ phase
transitions with $C=0.5,0.1,0.05,0$. The solid line denotes the
formation of the gapless CFL phase, whereas the dotted line does
the boundary of CFL quark matter.}
 \end{figure}

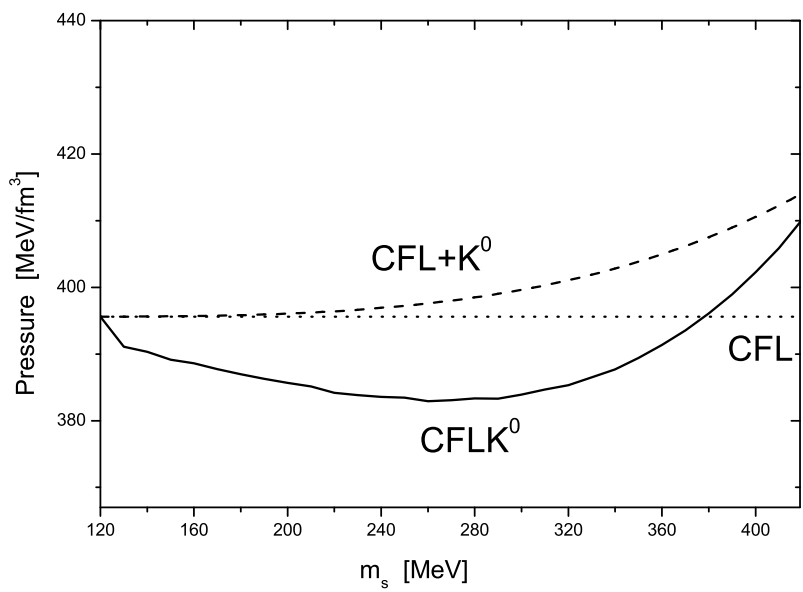
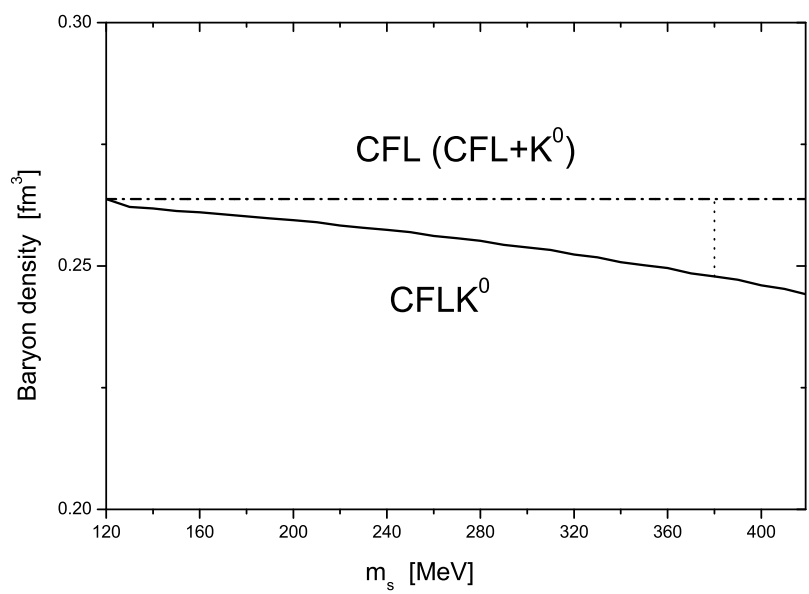
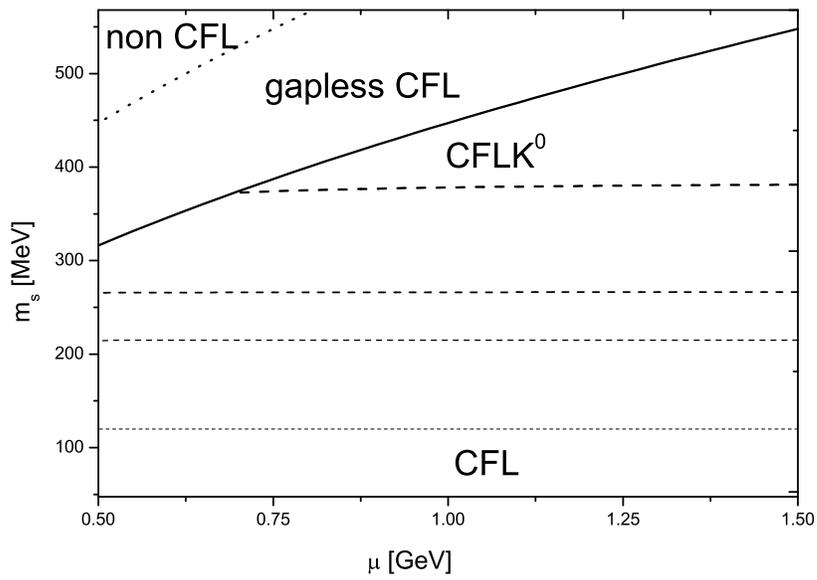
\end{document}